\documentstyle[twocolumn,aps]{revtex}% Two single-spaced columns
\newcommand{\beq}{\begin{equation}}
\newcommand{\eeq}{\end{equation}}
\newcommand{\half}{\mbox{$\textstyle \frac{1}{2}$} }
\newcommand{\ket}[1]{| #1 \rangle}
\newcommand{\bra}[1]{\langle #1 |}
\newcommand{\proj}[1]{\ket{#1}\! \bra{#1}}

\newcommand{\Tr}{{\rm Tr}}

\begin{document}
%\draft

\title {Purification of Noisy Entanglement and Faithful Teleportation via
Noisy Channels}

\author{Charles H. Bennett$^{(1)}$, Gilles Brassard$^{(2)}$,
Sandu Popescu$^{(3)}$, Benjamin Schumacher$^{(4)}$, John A.~Smolin$^{(5)}$,\\
and William K.~Wootters$^{(6)}$}

\address{(1) IBM Research Division, Yorktown Heights, NY 10598;
(2) D\'epartement IRO, Universit\'e de Montr\'eal, C.P.~6128,
Succursale centre--ville, Montr\'eal (Qu\'ebec), 
Canada H3C 3J7; (3) Physics Department, Tel Aviv University, Tel Aviv, Israel;
(4) Physics Department, Kenyon College, Gambier, OH 43022;
(5) Physics Department, University of California at Los Angeles, Los Angeles,
CA 90024; (6) Physics Department, Williams College, Williamstown, MA 01267}

\date{\today}
\maketitle

\begin{abstract}
{Two separated observers, by applying local operations to a
supply of not-too-impure entangled states ({\em e.g.}~singlets shared
through a noisy channel), can prepare a smaller number of
entangled pairs of arbitrarily high purity ({\em e.g.}~near-perfect singlets).
These can then be used to faithfully teleport unknown quantum states
from one observer to the other, thereby achieving faithful transmission
of quantum information through a noisy channel.  We give upper and lower bounds
on the yield $D(M)$ of pure singlets ($\ket{\Psi^-}$) distillable from mixed
states $M$, showing $D(M)>0$ if $\bra{\Psi^-}M\ket{\Psi^-}>\half$.}
\end{abstract}

\pacs{PACS numbers: 03.65.Bz, 42.50.Dv, 89.70.+c}
\narrowtext

The techniques of quantum teleportation~\cite{BBCJPW} and quantum data
compression~\cite{S95,JS94} exemplify a new goal of quantum
information theory, namely to understand the kind and quantity of channel
resources needed for the transmission of intact quantum states, rather
than classical information, from a sender to a receiver.

In this approach, the quantum source ${\cal S}$ is viewed as an
ensemble of pure states $\psi_i$, typically not all orthogonal, emitted
with known probabilities $p_i$. Transmission of quantum information
through a channel is considered successful if the channel outputs closely
approximate the inputs as quantum states.  Because non-orthogonal
states in principle cannot be observed without disturbing them, their
faithful transmission requires that the entire transmission processes be
carried out by a physical apparatus that functions obliviously, that is,
without knowing or learning which $\psi_i$ are passing through.

Just as classical data compression techniques allow data from a classical
source to be faithfully transmitted using a number of bits per       
signal asymptotically approaching the source's Shannon entropy,
$-\sum_ip_i\log_2p_i$, quantum data compression~\cite{S95,JS94} allows
quantum data to
be transmitted,  with asymptotically perfect fidelity, using a number of
2-state quantum systems or {\em qubits\/} ({\em e.g.}~spin-\half particles)
asymptotically approaching the source's von Neumann entropy
 \beq
S(\rho)=-\Tr\rho\log_2\rho, \hspace{3mm}\mbox{ where  }
\rho=\sum_ip_i\proj{\psi_i}.
 \eeq

Quantum teleportation achieves the goal of faithful transmission in a
different way, by substituting  classical communication and prior
entanglement for a direct quantum channel.  Using teleportation, an
arbitrary unknown qubit can be
faithfully transmitted via a pair of maximally-entangled qubits
({\em e.g.}~two spin-\half particles in a pure singlet state)
previously shared between sender and receiver, and a $2$-bit
classical message from the sender to the receiver.

Both quantum data compression and teleportation require a noiseless
quantum channel---in the former case for the direct quantum transmission
and in the latter for sharing the entangled particles---yet available channels
are typically noisy.  Since quantum information cannot be cloned~\cite{WZ}, it
would perhaps
appear impossible to use redundancy in the usual way to correct errors.
Nevertheless, quantum error-correcting codes have recently been
discovered~\cite{Shor95} which operate in a subtler way,
essentially by embedding the quantum information to be protected
in a subspace so oriented in a larger Hilbert space as to leak
little or no information to the environment, within a given noise model.
We describe another approach, in which the noisy channel is not used
to transmit the source states
directly, but rather to share entangled pairs ({\em e.g.}~singlets) for
use in teleportation.  But before they can be used to teleport reliably, the
entangled pairs must be purified---converted to almost-perfectly
entangled states from the mixed entangled states that result from
transmission through the noisy channel.  We show below how the
two observers can accomplish this purification, by performing local
unitary operations and measurements on the shared entangled pairs,
coordinating their actions through classical messages, and sacrificing
some of the entangled pairs to increase the purity of the remaining
ones.  Once this is done, the resulting
almost perfectly pure, almost perfectly entangled pairs can be used, in
conjunction with classical messages, to teleport the unknown quantum
states $\psi_i$ from sender to receiver with high fidelity.  The overall
result is to simulate a noiseless quantum channel by a noisy one,
supplemented by local actions and classical communication.

Let $M$ be a general mixed state of two spin-\half particles, from which
we wish to distill some pure entanglement.  The state $M$ could result,
for example, when one or both members of an initially pure singlet state
\mbox{$\Psi^-=(\uparrow\downarrow-\downarrow\uparrow)/\sqrt{2}$}
are transmitted through a noisy channel to two separated observers, whom
we shall call Alice and Bob.  The purity of $M$ can be
conveniently expressed by its fidelity~\cite{S95}
 \beq
F = \bra{\Psi^-}M\ket{\Psi^-}
 \label{purity}\eeq
relative to a perfect singlet.  Though
nonlocally defined, the purity $F$ can be computed from the probability
$P_{\parallel}$ of obtaining parallel outcomes if the two spins are
measured locally along the same random axis: one finds that
\mbox{$F = 1-3P_{\parallel}/2$.}

The recovery of entanglement from $M$ is best understood in the special
case that $M$ is already a pure state of the two particles,
$M=\proj{\Upsilon}$ for some $\Upsilon$. The quantity of entanglement,
$E(\Upsilon)$, in such a pure state is naturally defined by the von
Neumann entropy of the reduced density matrix of either particle
considered separately:
 \beq
E(\Upsilon)=S(\rho_A)=S(\rho_B),
\eeq\label{ee}\noindent
where $\rho_A=\Tr_B(\proj{\Upsilon})$, and similarly for $\rho_B$.  For
pure states, this entanglement can be efficiently concentrated into
singlets by the methods of \cite{BBPS95}, which use local operations
and classical communication to transform $n$ input states $\Upsilon$
into $m$ singlets with a yield $m/n$ approaching $E(\Upsilon)$ as
$n\rightarrow\infty$. Conversely, given $n$ shared singlets, local
actions and classical communication suffice to prepare $m$ arbitrarily
good copies of $\Upsilon$ with a yield $m/n$ approaching $1/E(\Upsilon)$
as $n\rightarrow\infty$.

Returning now to the problem of obtaining singlets from mixed states,
the first step in our purification protocol is to have
Alice and Bob perform a {\em random bilateral rotation\/} on each shared
pair, choosing a random SU(2) rotation independently for each pair and
applying it locally to both members of the pair (the same result could also
be achieved by choosing from a finite set of rotations $\{B_x, B_y, B_z, I\}$
defined below).  This transforms the
initial general two-spin mixed state $M$ into a rotationally symmetric
mixture
 \beq \begin{array}{rrrr}
 W_F  =  &         F\!\cdot\!\proj{\Psi^-}   & + &
                                  \frac{1\!-\!F}{3}\proj{\Psi^+} \\[1 mm]
      +  &  \frac{1\!-\!F}{3}\proj{\Phi^+}   & + &
\frac{1\!-\!F}{3}\proj{\Phi^-}
 \label{W}\end{array}
 \eeq of the singlet state $\Psi^-$ and the three triplet states
$\Psi^+=(\uparrow\downarrow+\downarrow\uparrow)/\sqrt{2}$ and
$\Phi^\pm=(\uparrow\uparrow\pm\downarrow\downarrow)/\sqrt{2}$. Because
of the singlet's invariance under bilateral rotations, the symmetrized
state $W_F$, which we shall call a Werner state~\cite{Werner} of purity
$F$, has the same $F$ as the initial mixed state $M$ from which it was
derived.

At this point it should be recalled that two mixed states having the
same density matrix are physically indistinguishable, even though they
may have had different preparations. Therefore, subsequent steps in the
purification can be carried out without regard to any properties of the
original mixed state $M$, or of the noisy channel(s) that may have
generated it, except for the purity $F$.

Mixtures of the four states $\Psi^\pm$ and $\Phi^\pm$---known
as the four Bell states---are
particularly easy to analyze because the Bell states transform simply
under several kinds of local unitary operations.  Besides the random
bilateral rotation already described, several other local
operations will be used in entanglement purification.

\smallskip\noindent$\bullet$ Unilateral Pauli rotations (that is, rotations
by $\pi$ radians
about the $x$, $y$, or $z$ axis) of {\em one\/} particle in an
entangled pair.  These operations map the Bell states onto one another
in a 1:1 pairwise fashion, leaving no state unchanged; thus $\sigma_x$
maps $\Psi^\pm \leftrightarrow \Phi^\pm$; $\sigma_z$ maps
$\Psi^\pm \leftrightarrow \Psi^\mp$ and $\Phi^\pm \leftrightarrow \Phi^\mp$,
while $\sigma_y$ maps $\Psi^\pm \leftrightarrow \Phi^\mp$.  We ignore overall
phase changes because they do not affect our arguments.

\smallskip\noindent$\bullet$ Bilateral $\pi/2$ rotations $B_x$, $B_y$, and
$B_z$ of {\em both\/} particles in a pair
about the $x$, $y$, or $z$ axis respectively.  Each of these operations
leaves the
singlet state and a different one of the triplets invariant,
interchanging the other two triplets, with $B_x$ mapping
$\Phi^+\leftrightarrow\Psi^+$, $B_y$ mapping
$\Phi^-\leftrightarrow\Psi^+$,
and $B_z$ mapping $\Phi^+\leftrightarrow\Phi^-$. Again we omit phases.

\smallskip\noindent$\bullet$ The quantum XOR or controlled-NOT
operation~\cite{Deutsch88}
performed bilaterally by both observers on corresponding members of two
shared pairs.  The {\em unilateral\/} quantum-XOR is an operation on two
qubits held by the same observer which conditionally flips the second
or ``target'' spin if the first or ``source'' spin is up, and
does nothing otherwise.  As a unitary operator it is expressed
 \beq \begin{array}{ccl}
U_{XOR}& =  & \ket{\uparrow_S\uparrow_T}\bra{\uparrow_S\downarrow_T}+
         \ket{\uparrow_S\downarrow_T}\bra{\uparrow_S\uparrow_T}  \\[1 mm]
    &   + &\ket{\downarrow_S\downarrow_T}\bra{\downarrow_S\downarrow_T} +
         \ket{\downarrow_S\uparrow_T}\bra{\downarrow_S\uparrow_T}.
 \end{array} \eeq The bilateral XOR (henceforth BXOR) operates in a
similar fashion on corresponding members of two pairs shared between
Alice and Bob: if Alice holds spins 1 and 3, and Bob holds
spins 2 and 4, a BXOR, with spins 1 and 2 as source and spins 3
and 4 as target would conditionally flip spin 3 if and only if spin 1
was up, while conditionally flipping spin 4 if and only if spin 2 was
up. A BXOR on two $\Phi^+$ states leaves them both
invariant. The result of
applying BXOR to other combinations of Bell states is shown below,
omitting phases.
 \beq\begin{array}{cc|cl}
\mbox{Before} && \mbox{After}& \mbox{(n.c. = no change)} \\
\hline
\mbox{Source}&\mbox{Target} & \mbox{Source}&\mbox{Target} \\
\hline
\Phi^\pm& \Phi^+ & \mbox{n.c.}&\mbox{n.c.} \\
\Psi^\pm& \Phi^+ & \mbox{n.c.}& \Psi^+ \\
\Psi^\pm& \Psi^+ & \mbox{n.c.}& \Phi^+ \\
\Phi^\pm& \Psi^+ & \mbox{n.c.}&\mbox{n.c.} \\
\Phi^\pm& \Phi^- & \Phi^\mp   &\mbox{n.c.} \\
\Psi^\pm& \Phi^- & \Psi^\mp   & \Psi^- \\
\Psi^\pm& \Psi^- & \Psi^\mp   & \Phi^- \\
\Phi^\pm& \Psi^- & \Phi^\mp   &\mbox{n.c.} \\
 \end{array}\eeq\label{bxor}

\smallskip\noindent$\bullet$ Besides these unitary operations, Alice and Bob
perform one kind of measurement: measuring both spins
in a given pair along the $z$ spin axis. This reliably
distinguishes $\Psi$ states from $\Phi$ states, but cannot distinguish
$+$ from $-$ states.  Of course after the measurement has been performed,
the measured pair is no longer entangled.

\medskip We now show that, given two Werner pairs of fidelity $F>\half$,
Alice and Bob can
use local operations and two-way classical communication to obtain, with
probability greater than $\frac{1}{4}$, one Werner pair of fidelity
$F'>F$, where the $F'$ satisfies the recurrence relation
 \beq
F' = \frac{F^2+\frac{1}{9}(1\!-\!F)^2}{F^2+\frac{2}{3}F(1\!-\!F)
+\frac{5}{9}(1\!-\!F)^2}.\label{recur}
 \eeq To achieve this, the following protocol is used:

\smallskip\noindent A1. A unilateral $\sigma_y$ rotation is performed on
each of the two pairs,
converting them from mostly-$\Psi^-$ Werner states to the analogous
mostly-$\Phi^+$ states, ie states with a large component $F>\half$ of
$\Phi^+$ and equal components of the other three Bell states.

\smallskip\noindent A2. A  BXOR is performed on the two impure $\Phi^+$
states, after which the target pair is locally measured along the $z$ axis.
If the target pair's $z$ spins come out parallel, as they would if both
inputs were true $\Phi^+$ states, the unmeasured source pair is kept;
otherwise it is discarded.

\smallskip\noindent A3. If the source pair has been kept, it is converted
back to a mostly-$\Psi^-$ state by a unilateral $\sigma_y$ rotation,
then made rotationally
symmetric by a random bilateral rotation (cf.~\cite{M95}).

\medskip Because $F'(F)$ is continuous and exceeds $F$ over the entire range
$\half < F < 1$, iteration of the above protocol can distill Werner states of
arbitrarily high purity $F_{\rm out}<1$ from a supply of input mixed
states $M$ of any purity $F_{\rm in}>\half$.  The yield
(purified output pairs per impure input pair) is rather poor, and tends to zero
in the limit $F_{\rm out}\rightarrow 1$; but by BXORing a variable
number $k(F)\approx 1/\sqrt{1\!-\!F}$ of source pairs, rather than 1,
into each target pair before measuring it, the yield can be increased
and made to approach a positive limit as $F_{\rm out}\rightarrow 1$.
(For this choice of $k$, to lowest order in $1\!-\!F$, the iteration
formula for purity agrees with Eq.~(\ref{recur}): $F'(F) = 1 -
\frac{2}{3}(1\!-\!F)$.  The expected fraction of the pairs discarded at
each step, also to lowest order, is $\frac{2}{3}(1\!-\!F)^\frac{1}{2}$.
One thus obtains a non-zero yield as
$F_{\rm out}\!\rightarrow\!1$.)
We do not give the asymptotic yield from this method,
because a higher yield can be obtained by combining it with another
method to be described below, which uses a supply of previously purified
$\Phi^+$
pairs in the manner of a breeder reactor, consuming some in order to
produce more than the number consumed.

The basic step (a ``BXOR test'') used in this method consists of
bilaterally  XORing a subset of the impure pairs, used as sources, into
one of the pure $\Phi^+$ states, used as a target, followed by
measurement of the target.  Consulting the table above, we see that each
$\Psi^+$ or $\Psi^-$ source pair toggles the target between $\Phi^+$ and
$\Psi^+$, without affecting the source.  Thus a BXOR test, like a parity
check on classical data, tells Alice and Bob whether there are an even
or odd number of $\Psi$ states in the tested subset.  By performing a
number of BXOR tests, on different subsets of the original impure pairs,
all the $\Psi$ states can be found and corrected to $\Phi$ states.  A
similar procedure is then used to find all the $\Phi^-$ states and correct
them to the desired $\Phi^+$.  The full protocol is described below.

\smallskip\noindent B1. Alice and Bob start with $n$ impure pairs described
by any Bell-diagonal
density matrix $W$ with $S(W)<1$, and $n\!\cdot\!(S(W)+\delta)$ prepurified
$\Phi^+$ states,
prepared, for example, by the variable blocksize recurrence method
described above.
Here $\delta$ is a positive constant that can be allowed to approach 0 in
the limit of
large $n$.

\smallskip\noindent B2. Using the prepurified $\Phi^+$ pairs as targets,
Alice and Bob perform
BXOR tests on sufficiently many random subsets of the impure pairs to locate
all $\Psi$ states, with high probability, without distinguishing
$\Psi^+$ from $\Psi^-$.  Once found, the $\Psi^{\pm}$ are
converted respectively to $\Phi^{\pm}$ by
applying a unilateral $\sigma_y$ rotation to each of them.  The
impure pairs now consist only of $\Phi^+$ and $\Phi^-$ states.

\smallskip\noindent B3. Next Alice and Bob do a bilateral $B_y$
to convert the $\Phi^-$ states into
$\Psi^+$, while leaving the $\Phi^+$ states invariant.  This done, they
perform  BXOR tests on sufficiently many more random subsets to find all
the new $\Psi^+$ states with high probability.  Once found, these states
are corrected to the desired $\Phi^+$ form by unilateral
$\sigma_x$ rotations.

\medskip The number of BXOR tests per impure pair required to find all the
errors, with arbitrarily small chance of failure, approaches the entropy
of the impure pairs, $S(W)=-\Tr W\log_2W$, in the limit of large $n$.
This follows from these facts:

\smallskip\noindent$\bullet$ For any two distinct $n$-bit strings the
probability that they agree on the parities of $r$ independent random
subsets of their bits is $\leq 2^{-r}$~\cite{BBCM}.

\smallskip\noindent$\bullet$  The probability distribution $P_X$ over
$n$-bit strings $x$,
where $x$ represents the original sequence of $\Phi/\Psi$ values of the
impure pairs, receives almost all its weight from a set of ``typical''
strings containing  $N_1=2^{H(X)+O(\sqrt{n}\,)}$ members, where $H(X)$ is
the Shannon entropy of $P_X$.  Similarly, the
conditional distribution $P_{Y|X=x}$ of $n$-bit strings $y$,
representing the $+/-$ values of a sequence of impure pairs whose
$\Phi/\Psi$ sequence is $x$, receives almost all its weight from a set
of typical strings containing $N_2=2^{H(Y|X=x)+O(\sqrt{n}\,)}$ members.

\medskip Let $r_1$ BXOR tests be performed in the first round, whose goal is
to find $x$ uniquely.  The expected number of ``false positives''---strings
in the typical set, other than the correct $x$, which agree with it on
$r_1$ subset parities---is $\leq N_12^{-r_1}$.  Thus the
chance of a false positive becomes negligible when $r_1>\log_2N_1$.
Similarly the chance of a false positive in the second round after $r_2$
BXOR tests is negligible when $r_2>\log_2N_2$.  Combining these
results, and recalling that $\log_2(N_1 N_2)=nS(W)+O(\sqrt{n}\,)$, we obtain
the desired result, viz.~that asymptotically $S(W)$ BXOR tests per impure pair
suffice to find all the errors.

The breeding method has a yield $1\!-\!S(W)$, producing more pure pairs
than consumed if the mixed state's von Neumann entropy, $S(W)$, is less
than 1.  For Werner states the yield
 \beq
 1-S(W_F)=1\!+\!F\log_2\!F+(1\!\!-\!\!F)\log_2\!\frac{1\!-\!F}{3}
 \label{D0}
 \eeq
is positive for $F\!>\!0.8107$.

The use of prepurified pairs as targets simplifies analysis of the
protocol by avoiding back-action of the targets on the sources, but is
not strictly necessary.  Even without the prepurified pairs, using only
impure Bell-diagonal states $W$ as input, it is possible~\cite{BDSW95}
to design a sequence of BXORs and local rotations that eliminate
approximately half the candidates for $x$ or $y$ at each step, achieving
the same asymptotic yield $1\!-\!S(W)$ as the breeding method. This
non-breeding protocol requires only one-way classical communication,
allowing it to be used to protect quantum information from errors during
storage (cf~\cite{Shor95,BDSW95}) as well as during transmission.

We do not yet know the optimal asymptotic yield $D(M)$ of purified
singlets distillable from general mixed states $M$, nor even from Werner
states.  Fig.~1 compares the yields of several purification methods for
Werner states $W_F$ with an upper bound $E(W_F)$ given by
 \beq\begin{array}{ll}
E(W_F) =   H_2(\half +\!\sqrt{F(1\!-\!F)})&\mbox{\bf ~~if~~}F\!\!>\!\!1/2 \\
E(W_F) =   0                           &\mbox{\bf ~~if~~}F\!\!\le\!\!1/2. \\
 \end{array}\label{upper}\eeq\noindent Here
$H_2(x)=-x\log_2x-(1\!-\!x)\log_2(1\!-\!x)$ is the dyadic Shannon
entropy. This upper bound is based on the fact that $W_F$, for
$F\!>\!\half$, can be expressed as an equal mixture of eight pure states
 \beq \sqrt{F}\ket{\Psi^-}+\sqrt{\frac{1-F}{3}}
(\pm\ket{\Psi^+}\pm\ket{\Phi^-}\pm i\ket{\Phi^+}),
 \label{Smolinstates} \eeq each having entropy of entanglement equal to
the right side of eq.~\ref{upper}a, while for
$F\!\!\leq\!\!\half$, $W_F$ can be expressed as mixture of
unentangled product states $\uparrow\uparrow, \downarrow\downarrow,
\uparrow\downarrow,$ and $\downarrow\uparrow$.  In fact~\cite{BDSW95}
these are the least entangled ensembles realizing $W_F$; therefore,
$E(W_F)$ may be viewed as the Werner state's ``entanglement of
formation''---the asymptotic
number of singlets required to prepare one
$W_F$ by local actions.  Because expected entanglement cannot
be increased by local actions and classical communication~\cite{BDSW95},
a mixed state's distillable entanglement $D(M)$ cannot exceed its
entanglement of formation $E(M)$.

We have seen that $F\!=\!\half$ is a threshold below which Werner
states can be made from unentangled ingredients, and above which they
can be used as a starting material to make pure singlets.  This is
further grounds (cf also \cite{Popescu}) for regarding all Werner states
with $F>\half$ as nonlocal even though only those with
$F>(2+3\sqrt{2})/8\approx0.78$ violate the
Clauser-Horne-Shimony-Holt~\cite{CHSH} inequality.  Distillable
entanglement and entanglement of formation are two
alternative extensions of the definition of entanglement from pure to
mixed states, but for most mixed states $M$ we do not know the value of
either quantity, nor do we know an $M$ for which they provably differ.

We thank David DiVincenzo for valuable advice.

\begin{figure}
\caption{Log-log plot of entanglement distillable from Werner
states of purity $F$ by various methods, versus $F\!\!-\!\!\half$.
$D_0$~=~breeding method alone (eq.~\protect\ref{D0});  $\;\;D_R$~=~breeding
preceded
by recurrence method of eq.~\protect\ref{recur};
$D_M$~=~ breeding preceded by recurrence of
\protect\cite{M95};
$E$~=~entanglement
of formation, eq.~\protect\ref{upper}, an upper bound on
entanglement yield of any method.}
\end{figure}


\begin{thebibliography}{99}

\bibitem{BBCJPW} C.H.~Bennett, G.~Brassard, C.~Cr\'epeau, R.~Jozsa,
A. Peres, and W.K.~Wootters, {\em Phys. Rev. Lett.} {\bf 70}, 1895
(1993).

\bibitem{S95} B. Schumacher, {\em Phys. Rev. A\/} {\bf 51}, 2738 (1995).

\bibitem{JS94} R.~Jozsa and B.~Schumacher,
{\em J. Modern Optics\/} {\bf 41}, 2343 (1994).

\bibitem{WZ} W.K.~Wootters and W.H.Zurek, {\em Nature\/} {\bf 299},
802 (1982).

\bibitem{Shor95} P.~Shor {\em Phys.~Rev. A\/} {\bf 52,\/} R2493 (1995).

\bibitem{BBPS95} C.H.~Bennett, H.~Bernstein, S.~Popescu, B.~Schumacher,
``Concentrating Partial Entanglement by Local Operations'',
submitted to {\em Phys.~Rev.~A\/} (1995).

\bibitem{Werner} R.F.~Werner, {\em Phys. Rev. A\/} {\bf 40},
4277 (1989).

\bibitem{Deutsch88} D. Deutsch, {\em Proc. Roy. Soc. A}
{\bf 425}, 73 (1989).

\bibitem{M95} C.~Macchiavello (private communication 1995)
found an improved recurrence using $B_x, \sigma_y$ in place of our step A3.

\bibitem{BBCM} C.H.~Bennett, G.~Brassard, C.~Cr\'epeau, and U.M.~Maurer,
``Generalized Privacy Amplification'', to appear Nov.~1995 in {\em IEEE
Trans. on Information Theory}.

\bibitem{BDSW95} C.H.~Bennett, D.~DiVincenzo, J.A.~Smolin, and W.K.~Wootters
``Mixed State Entanglement and Quantum Error Correcting Codes'', preprint
(1995).

\bibitem{Popescu} S.~Popescu, {\em Phys. Rev. Lett.\/}
{\bf 72}, 797 (1994).

\bibitem{CHSH} J.F.~Clauser, M.A.~Horne, A.~Shimony, R.A.~Holt,
{\em Phys.~Rev.~Lett.\/} {\bf 23}, 880 (1980).

\end{thebibliography}
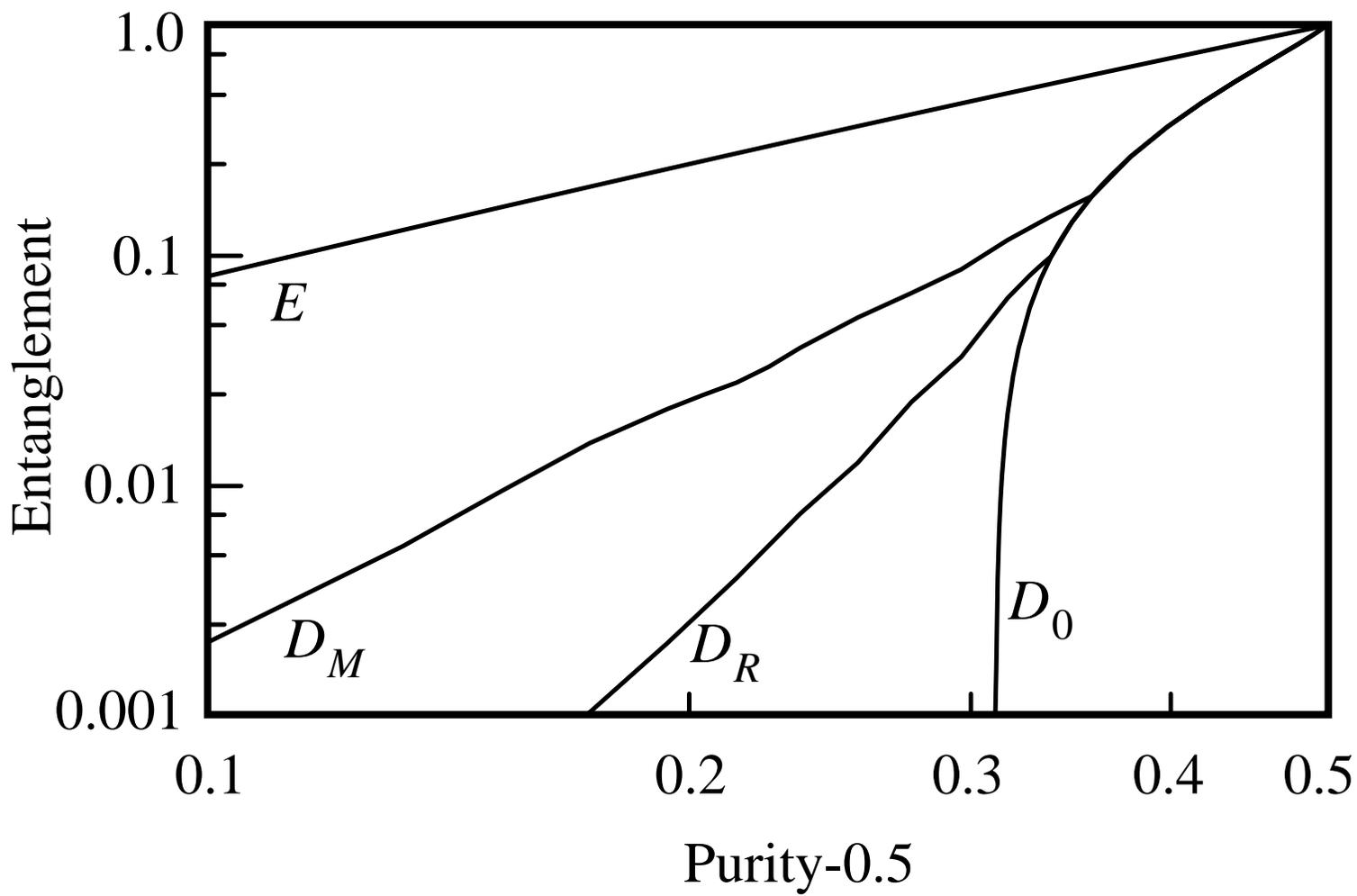
\end{document}